\documentclass[aps, prl, reprint, floatfix, amsmath, amssymb, showpacs]{revtex4-1}
\usepackage[utf8]{inputenc}
\usepackage[T1]{fontenc}
\usepackage{amsmath,amssymb}
\usepackage{graphicx}
\usepackage{tikz}
\usepackage{color}
\usepackage{microtype}
\usepackage{float}

\begin{document}

\title{Knots as Topological Order Parameter for Semi-Flexible Polymers}
\author{Martin Marenz}
\email{martin.marenz@itp.uni-leipzig.de}
\author{Wolfhard Janke}
\email{wolfhard.janke@itp.uni-leipzig.de}
\affiliation{Institut f\"ur Theoretische Physik, Universit\"at Leipzig, Postfach 100\,920, D-04009 Leipzig, Germany}


\begin{abstract}
  Using a combination of the replica-exchange Monte Carlo
  algorithm and the multicanonical method, we investigate the influence of
  bending stiffness on the conformational phases of a bead-stick homopolymer
  model and present the pseudo-phase diagram for the complete range of
  semi-flexible polymers, from flexible to stiff. Although a simple
  model, we observe a rich variety of conformational phases, reminiscent of
  conformations observed for synthetic polymers or biopolymers. Changing the internal
  bending stiffness, the model exhibits different pseudo phases like bent,
  hairpin or toroidal. In particular, we find thermodynamically stable knots
  and transitions into these knotted phases with a clear phase coexistence, but
  almost no change in the mean total energy.
\end{abstract}

\maketitle


Since the first simulation of knotted polymers in
1975~\cite{Frank-Kamenetskii1975}, the occurrence and behavior of knots in
polymers has been studied in various contexts. Scanning through protein data
bases has revealed that several proteins form knots~\cite{Mansfield1994,
Taylor2000, Lua2006}. In particular, Virnau et al.~\cite{Virnau2006} have
reviewed the whole Protein Data Bank (http://www.pdb.org~\cite{Berman2000}) and
identified 36 different proteins forming relatively simple knots, none of which
features more than five crossings -- somehow evolution tries to avoid knotted
proteins~\cite{Reith2015}. On the other hand, flexible polymers form much more
complicated knots, which occur by chance in the
swollen~\cite{Koniaris1991b,Deguchi1997,Virnau2005a} and in the collapsed
phase~\cite{Lua2004,Virnau2005a}. Essential for the existence of these two
phases is the excluded volume and attraction of the monomers. Already lattice
polymer simulations~\cite{Bastolla1997,Krawczyk2010} show that models
integrating self-avoidance and attraction exhibit a swollen, a globular,
and a frozen phase. In this work we go a step ahead and investigate the
knottedness of semi-flexible bead-stick polymers. There exists already some
works concerning the more complex phase space of polymer models that
incorporate bending stiffness~\cite{Noguchi1997}. The most comprehensive study
uses a model for semi-flexible polymers of bead-spring type with finitely
extensible nonlinear elastic (FENE) bonds~\cite{Seaton2013}. By varying the
bending stiffness, these models are able to mimic a large class of polymers,
exhibiting, for instance, bent, hairpin or toroidal conformations.

Nevertheless, none of the former work considered the knottedness of the polymer
over the full bending stiffness range, which we will discuss in this Letter. By
measuring the knot type we found pseudo phases with thermodynamically stable
knotted polymers. The knot type will be shown to be an ideal topological order
parameter to identify the knotting transition and, moreover, the behavior at
the transition from an unknotted to a knotted pseudo phase turns out to be
surprisingly different from all other pseudo-phase transitions of the
bead-stick polymer in that it does not fit into the common classification
scheme of first- and second-order phase transitions.

To model a coarse-grained polymer with an adjustable stiffness, we use a
modified version of the bead-stick model of
Refs.~\cite{Stillinger1993,Irback1997,Bachmann2005}, which neglects the
hydrophobicity of the monomers. The model then consists of $N$ identical
monomers connected by bonds with length one. Non-adjacent monomers interact via
the Lennard-Jones potential
\begin{align}
  \label{form:LJ}
 E_{\textrm{LJ}} = 4 \epsilon
 \sum\limits_{i=1}^{N-2} \sum\limits_{j=i+2}^{N} \left[ \left(
  \frac{\sigma}{r_{ij}}\right)^{12} - \left( \frac{\sigma}{r_{ij}} \right)^{6}
  \right],
\end{align}
where $r_{ij}$ is the distance between two monomers. The parameters $\epsilon$
and $\sigma$ are set to one for the rest of this work, i.e., energies are
measured in units of $\epsilon$ and lengths in units of $\sigma$.  The
stiffness is modeled through the cosine potential adopted from the well-known
discretized worm-like chain model~\cite{Rubinstein2003} and defined by
\begin{align}
  \label{form:Bend}
 E_{\textrm{bend}} = \sum\limits_{i=1}^{N-2} \left( 1 - \cos \theta_i \right),
\end{align}
where $\theta_i$ represents the angle between adjacent bonds.  The complete
Hamiltonian is then given by $ E = E_{\textrm{LJ}} + \kappa E_{\textrm{bend}}$,
where $\kappa$ parametrizes the strength of the bending term compared to
the Lennard-Jones potential.

Topological barriers between the knotted phases forced us to apply relatively
complex Monte Carlo algorithms in order to obtain reliable results. To
simulate the system in the complete $(T,\kappa)$-plane, we used two
complementary Monte Carlo algorithms. The first is a combination of the
parallel multicanonical method~\cite{Zierenberg2013,Berg1991,*Berg1992} and a
one-dimensional replica  exchange~\cite{Hukushima1995} in the $\kappa$
direction (PMUCA+RE). The second is a two-dimensional version of the
replica-exchange method (2D-RE), which simulates the system in $T$ and $\kappa$
direction in parallel. By means of a two-dimensional weighted histogram
analysis algorithm~\cite{Ferrenberg1988, *Ferrenberg1989}, we are able to
calculate the canonical mean values for every point in the $(T,\kappa)$-plane
within the simulated parameter ranges. To generate well equilibrated results it was
necessary to apply quite intricate bridge-end and double-bridging moves that
respect the fixed bond-length constraint, besides the common crank-shaft,
spherical-rotation and pivot moves.  The results of both methods, PMUCA+RE and
2D-RE, are in good agreement with each other.

To determine the structural phases we measured the total energy $\left\langle E
\right\rangle$, both sub-energies $\left\langle E_{\textrm{LJ}} \right\rangle$
and $\left\langle E_{\textrm{bend}} \right\rangle$, the squared end-to-end
distance $\left\langle R^2_{\textrm{ee}} \right\rangle$, the squared radius of
gyration $\left\langle R_g^2 \right\rangle$, and the eigenvalues of the
gyration tensor $\left\langle \lambda_1 \right\rangle, \left\langle \lambda_2
\right\rangle,\left\langle \lambda_3 \right\rangle$. The derivatives with
respect to temperature of these observables mark the locations of the different
pseudo-phase transitions, which should not be mistaken with phase transitions
in the thermodynamic limit ($N \to \infty$). The different observables give
slightly different results for finite systems, so that the transitions are
smeared out. Additionally, we performed microcanonical analyses as a
complementary approach to identify the different pseudo-phase transitions.
These are based on the microcanonical entropy, $S(E) = k_{\textrm{B}} \ln
\Omega (E)$, with $\Omega(E)$ being the density of states.  In the
microcanonical framework a peak in the derivative of the microcanonical inverse
temperature $ d\beta_{\textrm{micro}}(E) / dE$, which is itself the derivative
of the entropy, $\beta_{\textrm{micro}}(E) = dS(E)/dE$, corresponds to a
pseudo-phase transition. A detailed description of the framework can be found
in Ref.~\cite{Junghans2006}. Although it is expected that all but the collapse
and freezing transitions vanish in the thermodynamic limit, they still
determine the structural behavior of short and midrange polymers and are
important for understanding the behavior of mesoscopic systems.

\begin{figure}
    
  \begin{minipage}{0.45\textwidth}
    \includegraphics[width=\textwidth,trim=0 50 0 0,clip=true]{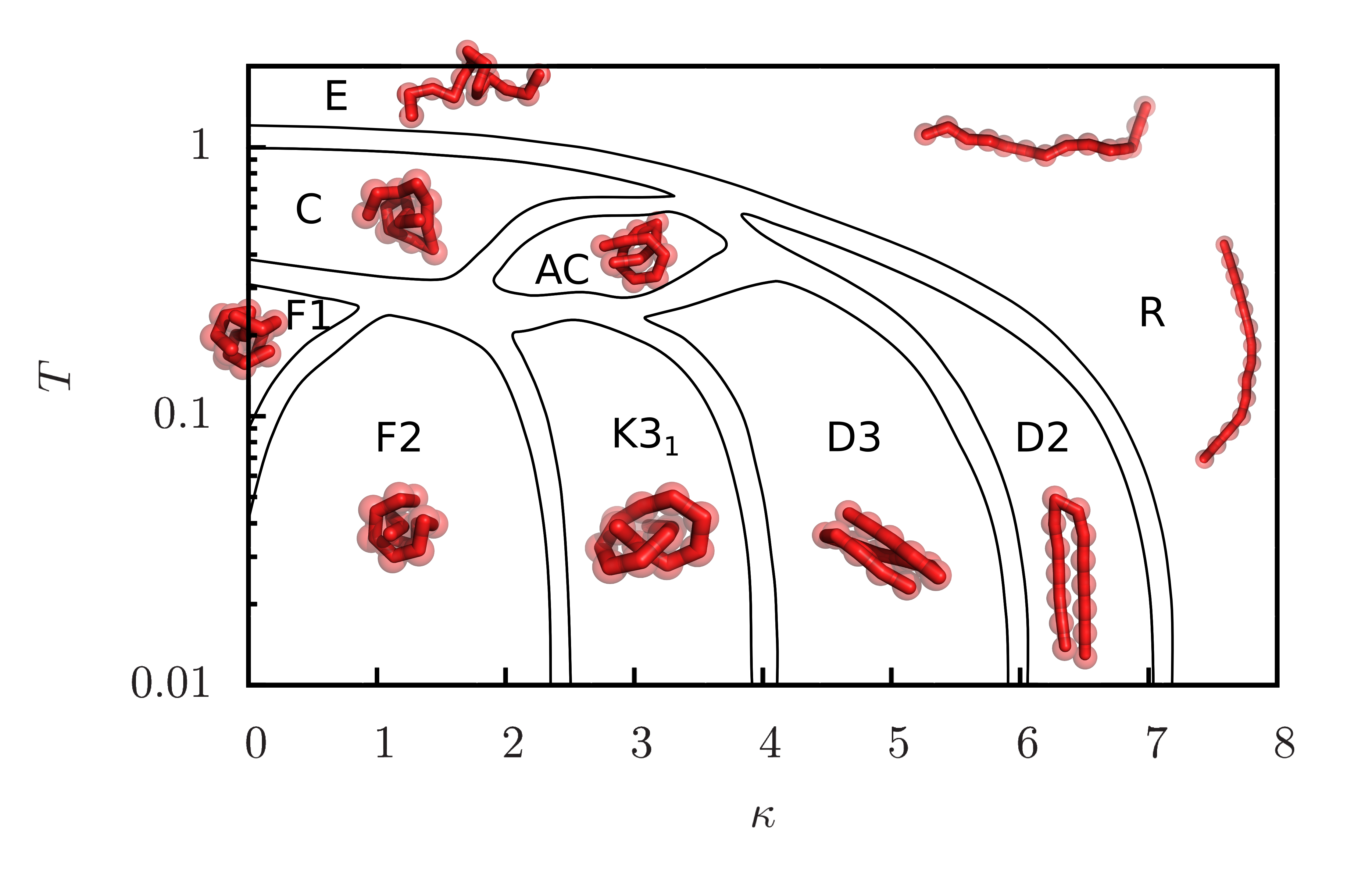}
  \end{minipage}
  \parbox{0.03\textwidth}{(a)}
  \parbox{0.44\textwidth}{\hfill}
  \begin{minipage}{0.45\textwidth}
    \hspace{-0.25cm}
    \includegraphics[width=0.94\textwidth,trim=10 20 0 0,clip=true]{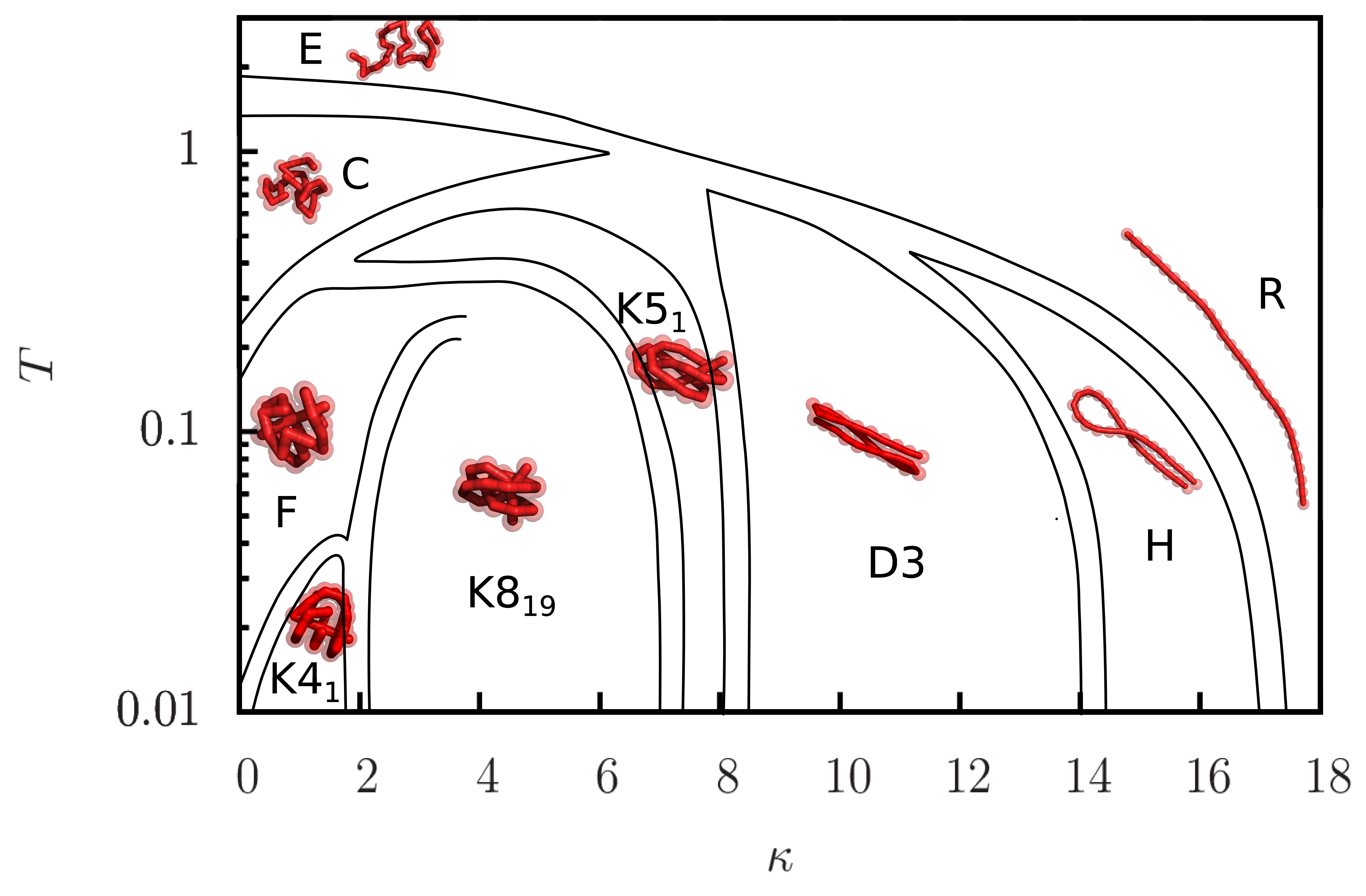}
  \end{minipage}
  \parbox{0.03\textwidth}{(b)}
  \parbox{0.44\textwidth}{\hfill}

  \caption{\label{fig:Phases} (color online) The black lines sum up all signals of the
     thermal derivative of all measured observables for (a) 14 monomers and (b)
     28 monomers. The pseudo phases are labeled as follows: E - elongated, R -
     rod-like, C - collapsed (AC is an artifact of the small chain length,
     both termini of the polymer are aligned), F - frozen, K$C_n$ - knotted
     phase with the corresponding knot type, D$N$ - $(N-1)$ times bent
     polymers, H - hairpin. We omitted some sub-phases of D$N$, K$8_{19}$, and
   F, which do not change the overall picture.}
\end{figure}

The stiffness-dependent behavior of the bead-stick model
[Eqs.\,(\ref{form:LJ}),\,(\ref{form:Bend})] for chains with $N=14 \textrm{ and
}28$ monomers is summed up in the pseudo-phase diagrams shown in
Fig.~\ref{fig:Phases}, which are constructed from the surface plots of all
measured thermal derivatives and the results of the microcanonical analysis.
The black lines in Fig.~\ref{fig:Phases} mark the thermally most active regions
and represent the location of the pseudo-phase transitions. For high
temperatures, where the system is entropy dominated and the polymer resembles a
discretized worm-like chain, the conformations are either gaseous-like and
extended (E) or rod-like (R). When the temperature is lowered, the
Lennard-Jones energy becomes more important and the polymer collapses. One can
clearly distinguish two different regimes. For small $\kappa$ the polymer
behaves similarly to a flexible chain with collapsed (C) and frozen (F) pseudo
phases. This contrasts with larger $\kappa$, where the polymer undergoes a
first-order-like transition from the unstructured state (R) to states with
differently structured motifs ($\textrm{D}N$, H, K$C_n$). For clarity, we
omitted in Fig.~\ref{fig:Phases} some sub-phases, where the shape parameters
and the microcanonical analysis suggest additional pseudo-phase transitions
between differently shaped conformations of the same motif. Especially, the
frozen phase F subsumes many different crystal-like phases which differ only in
minor aspects.

To identify the knotted phases, we measured the knot type of the polymer, which
turns out to be an ideal order parameter. In principle, the knot type, denoted
by $C_n$, defines which smooth closed curves can be transformed into each other
by applying multiple Reidemeister moves. Practically, this means two knots are
not of the same type if one of them cannot be deformed into the other without
cutting and rejoining the curve. The integer number $C$ counts the minimal
number of crossings for any projection on a plane and the subscript $n$
distinguishes topologically different knots with the same number of crossings.
A detailed exposition of mathematical knot theory can be found in
Ref.~\cite{Kauffmann1991}. Of course, an open polymer cannot satisfy the
mathematical definition of a knot, unless the termini are closed virtually. In
this work we applied a closure which is inspired by tying a real knot. Both
termini are connected by a straight line, which is enlarged in both directions
to a point far outside the polymer. These new termini are then connected
through a point which lies on a line perpendicular to the connecting line of
the termini and which is also located far away from the center of mass of the
polymer. We have tested our results with different other closures and obtained
almost identical results. To calculate $C_n$, we employed a technique described in
Ref.~\cite{Virnau2010}, which is based on a variant of the Alexander polynomial
$\Delta(t)$,
\begin{align} 
  \Delta_p(t) = | \Delta(t) \times \Delta(1/t) |,
\end{align}
evaluated at $t = -1.1$. $\Delta_p(t)$ inherits the ability to distinguish
different knot types from the Alexander polynomial. Strictly speaking, the
Alexander polynomial and likewise $\Delta_p(t)$ are not unique invariants. However,
for simpler knots both generate unique results, and they are able to distinguish
all knots found in this work.

\begin{figure}
  \begin{minipage}{0.45\textwidth}
  \hspace{-0.55cm}
    \includegraphics[width=\textwidth,trim=0 50 0 180,clip=true]{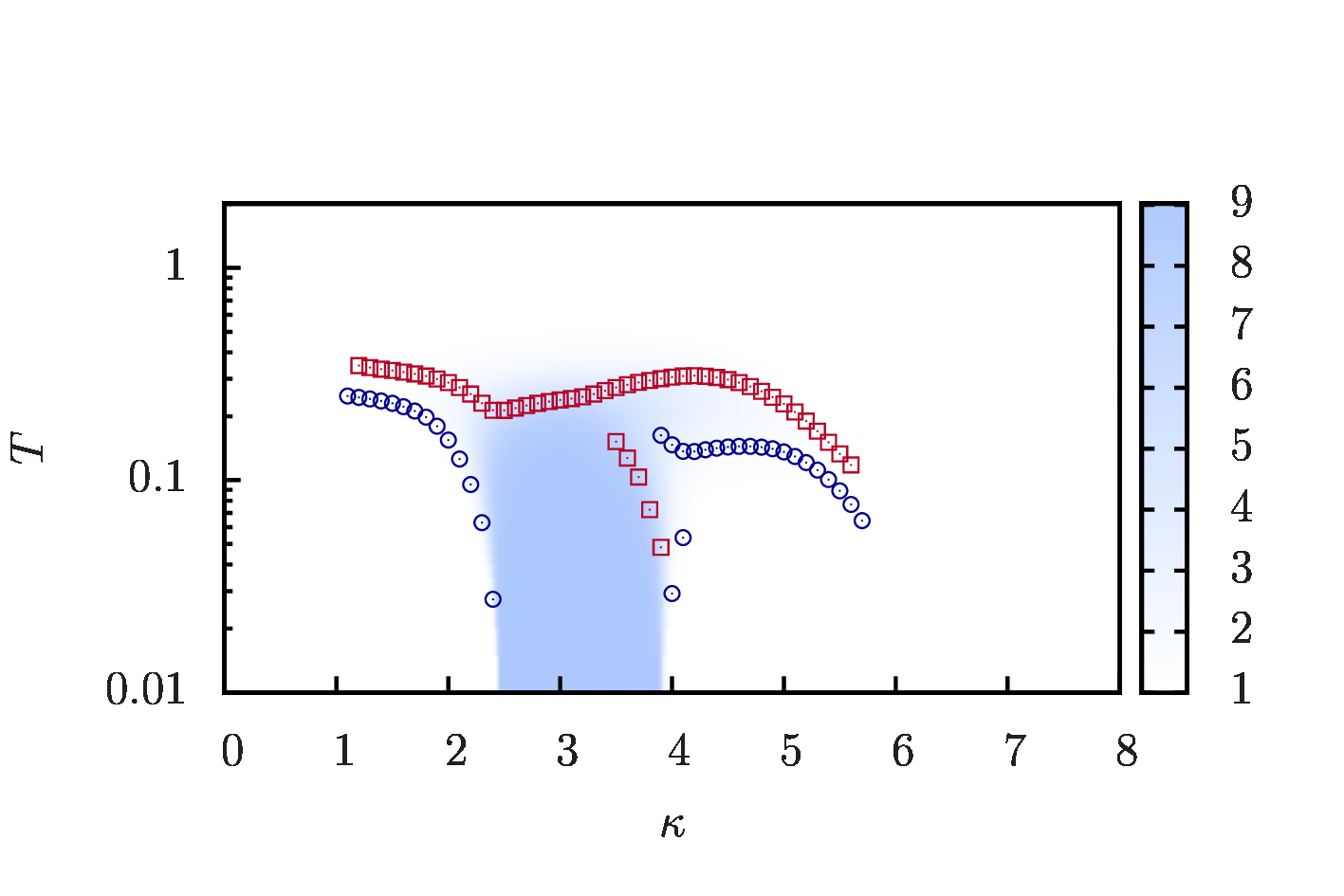}
  \end{minipage}
  \parbox{0.03\textwidth}{(a)}
  \parbox{0.44\textwidth}{\hfill}
  \begin{minipage}{0.45\textwidth}
    \includegraphics[width=\textwidth]{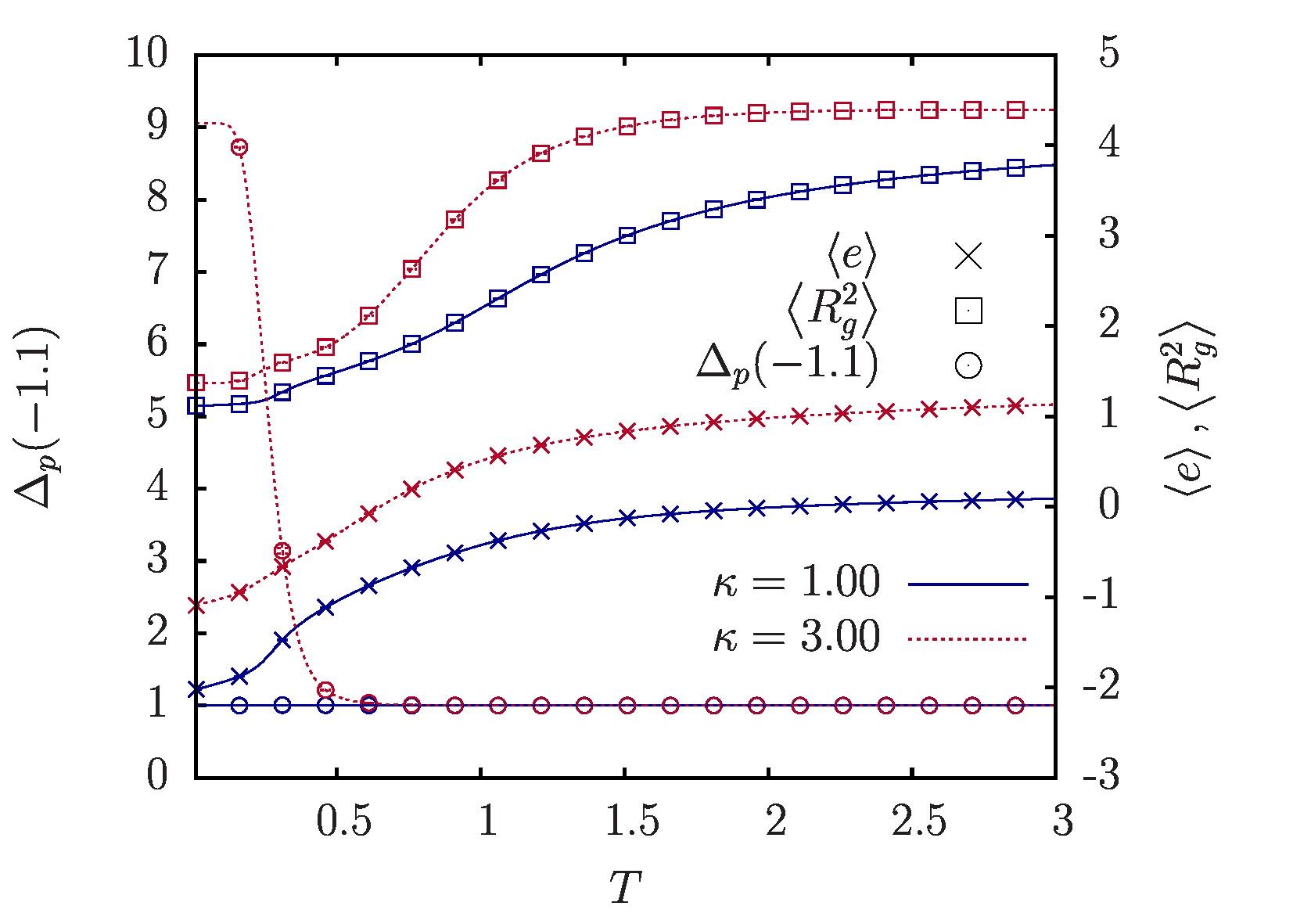}
  \end{minipage}
  \parbox{0.03\textwidth}{(b)}
  \parbox{0.44\textwidth}{\hfill}

  \caption[fragile]{\label{fig:N14_Knot}(color online) (a) Surface plot of
    $\Delta_p(-1.1)$ over the complete $(T, \kappa)$-plane for a 14mer, blue
    circles correspond to maxima and red squares to minima of $\frac{d}{dT}
    \Delta_p(-1.1)$. The blue regime,
    $\Delta_p(-1.1) = 9.0546 $, marks the $\textrm{K}3_1$ knot phase of the
    polymer. (b) Temperature profile  of  $\left\langle E/N \right\rangle$,
    $\left\langle R_g^2 \right\rangle$, and $\Delta_p(-1.1)$ at $\kappa = 1.00
    \textrm{ (blue solid lines) and } \kappa = 3.00 \textrm{ (red dashed
    lines)}$. In contrast to $\left\langle E/N \right\rangle \textrm{ and }
    \left\langle R_g^2 \right\rangle$, $\Delta_p(-1.1)$ differs significantly
    between the knotted ($\kappa = 3.00$) and unknotted ($\kappa=1.00$) phase.}
\end{figure}

In the knotted pseudo phases, marked with K in Fig.~\ref{fig:Phases}, the
probability to find the corresponding knot is almost one. This means
every measured polymer is of that knot type and implies that the knots are
thermodynamically stable. We want to emphasize that this is different from the
formerly investigated knots which form by chance in the swollen or collapsed
phase. This makes $\Delta_p(-1.1)$ a perfect topological
order parameter to distinguish knotted and unknotted phases, as is demonstrated in
Fig.~\ref{fig:N14_Knot}(a). In Fig.~\ref{fig:N14_Knot}(b) one can see that the
behavior of the mean total energy $\left\langle E \right\rangle$ and the squared radius of
gyration $ \left\langle R_g^2 \right\rangle $ is qualitatively very
different at the knotting transition $ \textrm{AC} \leftrightarrow \textrm{K}3_1
$ ($\kappa=3.00$) and the freezing transition $\textrm{C}
\leftrightarrow \textrm{F}2$ ($\kappa=1.00$). On the other hand, the knot
parameter $\Delta_p(-1.1)$ clearly signals the pseudo-phase transition and goes
from 1 (unknotted polymer) to 9.0546 ($3_1$ knot) only in the case of a
transition into a stable knot. As expected, the 28mer (and even more the
42mer~\footnote{Not shown here, because the details of the more complicated
pseudo-phase diagram do not contribute to the understanding of the basic
mechanism.}) exhibits a richer phase diagram with more complicated knot types,
see Fig. 1(b). However, the qualitative behavior at the phase boundaries turned
out to be very similar so that we will focus in the following on the 14mer.

\begin{figure}
  \hspace{-0.4cm}
  \begin{minipage}{0.45\textwidth}
    \includegraphics[width=0.9\textwidth]{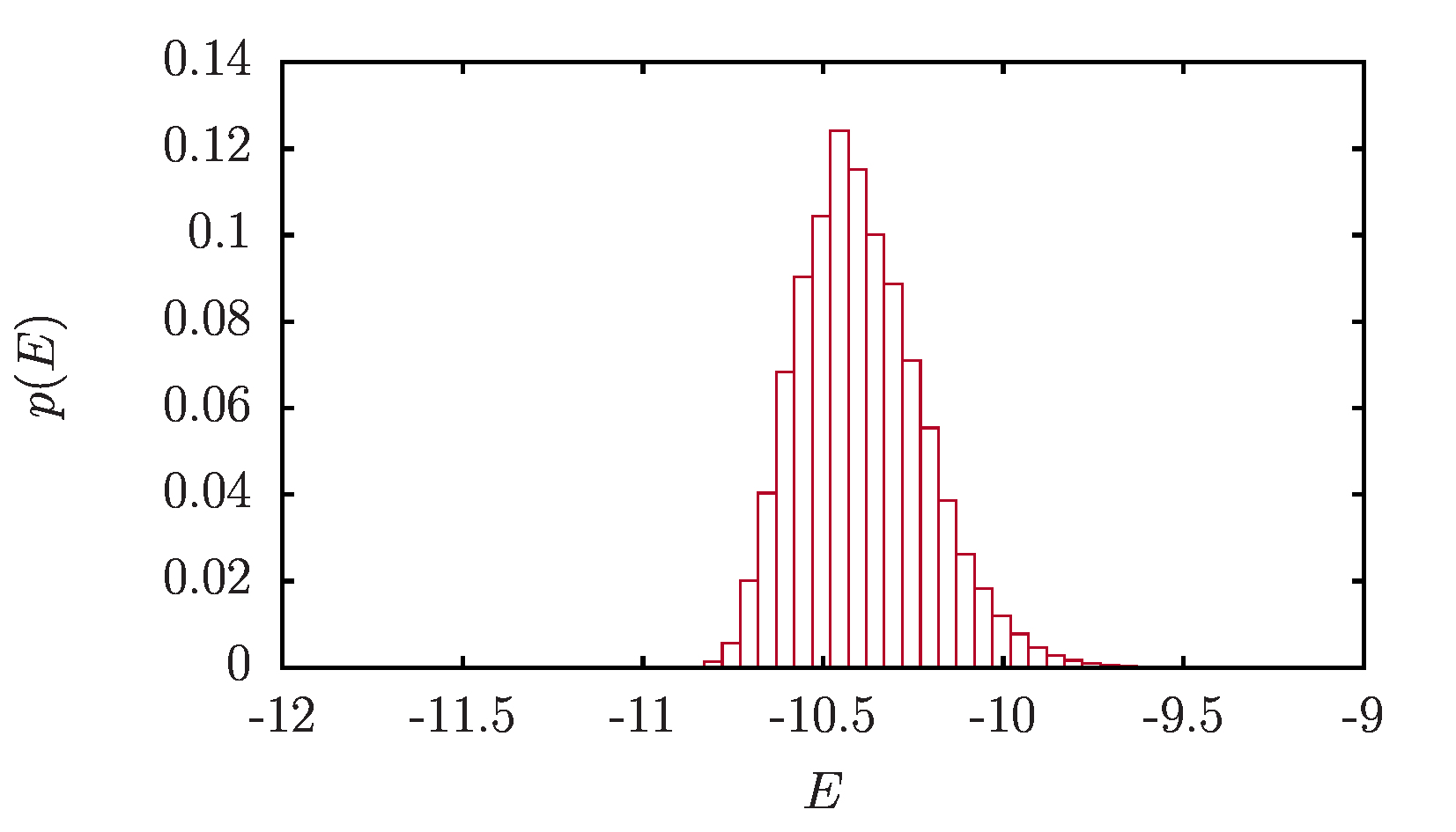}
  \end{minipage}\\
  \vspace{-0.1cm}
  \parbox{0.03\textwidth}{(a)}
  \vspace{-0.35cm}
  \parbox{0.44\textwidth}{\hfill}
  \begin{minipage}{0.45\textwidth}
    \includegraphics[width=0.9\textwidth, trim=0 10 0 10,clip=true]{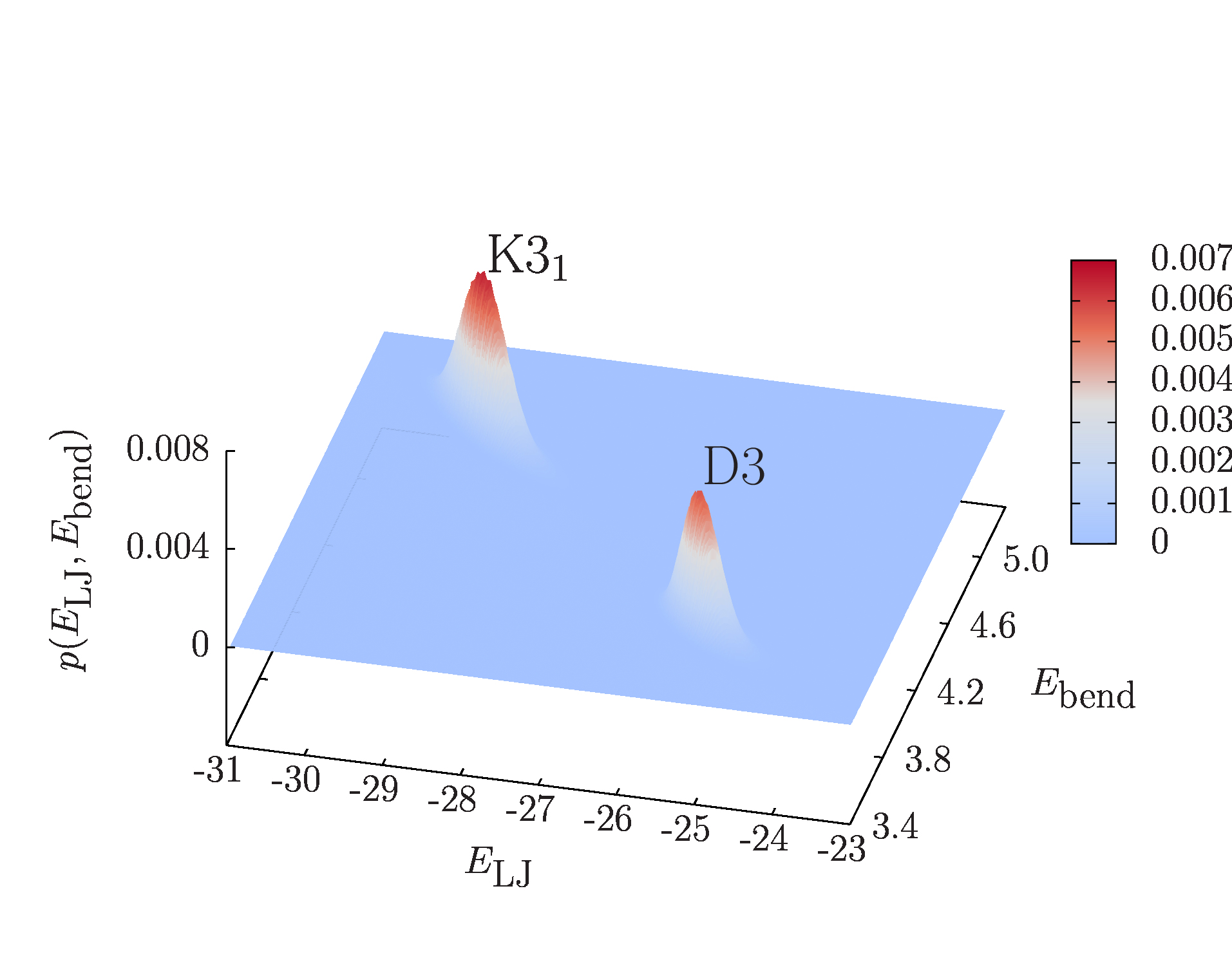}
  \end{minipage}
  \parbox{0.03\textwidth}{(b)}
  \parbox{0.44\textwidth}{\hfill}

  \caption{\label{fig:EnergyDistribution}
    (color online)
    Energy probability distribution of a 14mer at a point where $\frac{d}{dT}
    \Delta_p(-1.1)$ suggests a knotting transition ($\kappa = 3.9$ and
    $T=0.0483$). Plot (a) shows the one-dimensional probability distribution
    $p(E)$. In (b) the energy is split into the two sub-energies, which leads to
    a double peak in the two-dimensional probability distribution
    $p(E_{\textrm{LJ}}, E_{\textrm{bend}})$. The left peak corresponds to the
    knotted state $\textrm{K}3_1$ and the right one to the bent state
    $\textrm{D}3$. The one-dimensional probability distribution is exactly the
    projection of $p(E_{\textrm{LJ}}, E_{\textrm{bend}})$ along the line
    connecting the two peaks, thus the phase coexistence is perfectly hidden in
    $p(E)$.}
\end{figure}

\begin{figure}
  \includegraphics[width=0.45\textwidth, trim=0 0 0 0, clip=true]{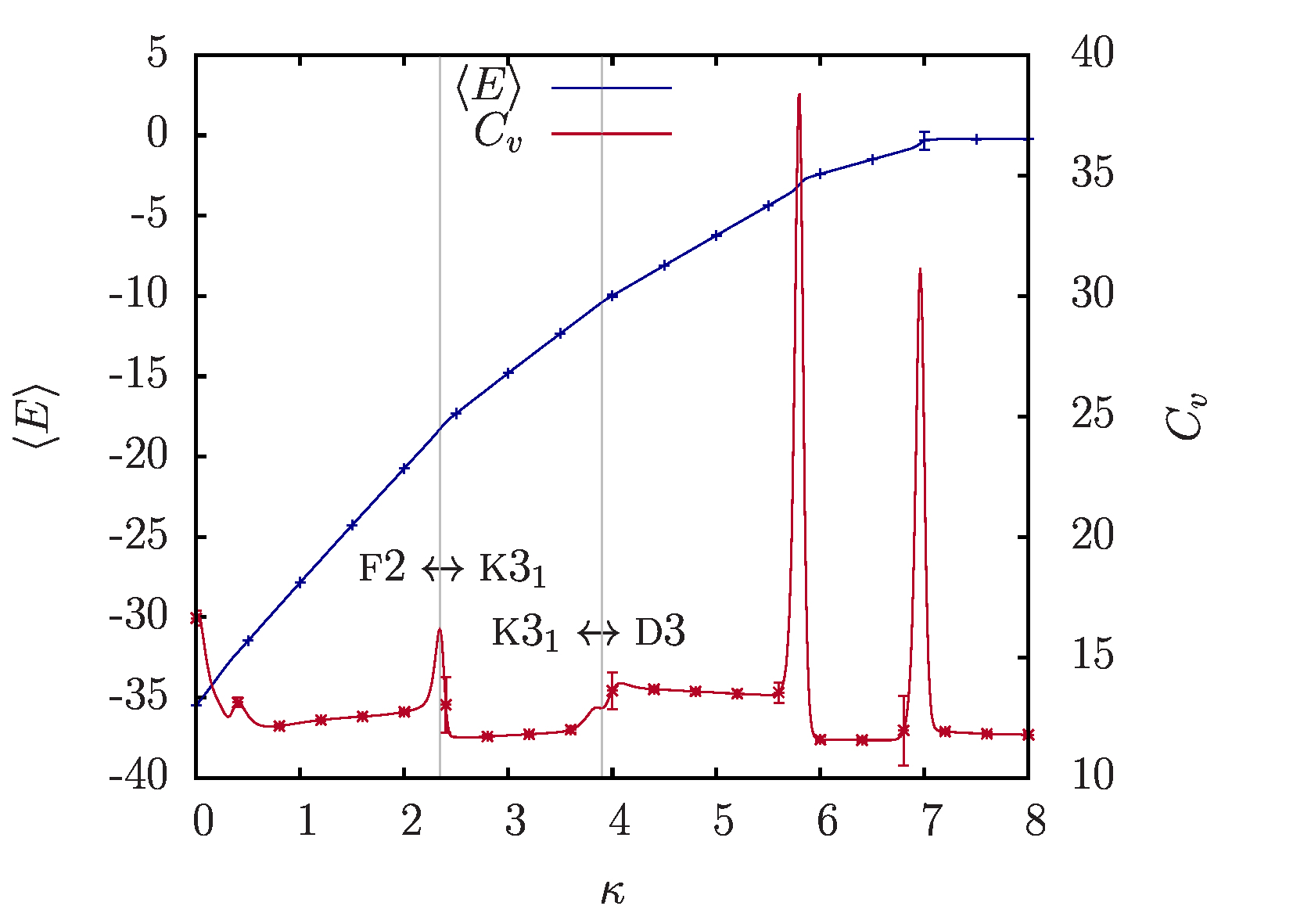}
  \caption{\label{fig:EnergyKnotTrans} (color online) Energy $\left\langle E
    \right\rangle$ and heat capacity $C_v$ of a 14mer at $T = 0.0483$ over the
    complete range of investigated $\kappa$ values. At both knotting
    transitions, $\kappa=2.7: \textrm{F}2 \leftrightarrow \textrm{K}3_1$ and
    $\kappa=3.9: \textrm{K}3_1 \leftrightarrow \textrm{D}3$, there is nearly no
    shift in the energy and therefore no signal in the heat capacity that is larger
    than the statistical error.}
\end{figure}

The knotting transitions from one structured state to another (e.g.,
$\textrm{K}3_1 \leftrightarrow \textrm{D}3$) are quite interesting. At first
glance, one could assume that they behave first-order-like, similar to other
solid-solid-like transitions at low temperatures. However, the microcanonical
analysis indicates a second-order-like behavior by a peak in
$d\beta_{\textrm{micro}}(E)/ dE$ which is smaller than zero (not shown here).
Likewise, the canonical probability distribution $p(E)$ does not exhibit a
double-peak structure, see Fig.~\ref{fig:EnergyDistribution}(a).  On the other
hand, the two-dimensional energy distribution $p(E_{\textrm{LJ}},
E_{\textrm{bend}})$ points to a phase coexistence. In
Fig.~\ref{fig:EnergyDistribution}(b) one can clearly identify two separate
peaks, one corresponding to the knotted phase and the other to the unknotted
phase. Surprisingly, both phases have almost identical mean total energy
$\left\langle E \right\rangle$ at the coexistence point, and there is almost no
signal in the total energy and heat capacity at the transition, see
Fig.~\ref{fig:EnergyKnotTrans}. We thus observe no latent heat where the
polymer undergoes the transition into the knotted phase. Rather, the
Lennard-Jones energy $E_{\textrm{LJ}}$ and the bending energy
$E_{\textrm{bend}}$ are transformed into each other~\footnote{This topological
 change also explains why we need PMUCA+RE or 2D-RE to overcome the
topological barrier.}.

This behavior changes if the polymer enters the knotted phase from an
unstructured conformation. For example, staying on the transition line between
$\textrm{K}3_1 \leftrightarrow \textrm{D}3$ for $N=14$ and going to higher
temperatures, the two peaks in $p \left( E_{\textrm{LJ}}, E_{\textrm{bend}}
\right)$ start to merge until they form a single peak at the transition between
$\textrm{AC} \leftrightarrow \textrm{K}3_1$, where the phase coexistence
vanishes. This behavior is similar for all other observed knotting
transitions.

The reason for the missing knotted phases in Ref.~\cite{Seaton2013} is not
rooted in this intricate behavior with ``concealed'' signals, however,
but may rather lie in the choice of the bond length $r_{b}$ and the minimum of
the Lennard-Jones potential $r_{\textrm{min}}$ which are identical
in~\cite{Seaton2013} but different by a factor of $1.22$ in this work. We have
tested a few parametrizations of bead-spring models and always observed
thermodynamically stable knots except for the special parametrizations used
in~\cite{Seaton2013}. The observation that the polymer minimizes its total
energy in bent conformations by maximizing the number of monomers located in
the Lennard-Jones minima of other monomers suggests the following conjecture:
if $r_b \approx r_{\textrm{min}}$ and the bonds are flexible enough, bent
conformations are energetically so strongly favored that knotted states become
unlikely.

In conclusion, we have investigated the complete stiffness dependent behavior
of a semi-flexible bead-stick homopolymer. Besides the conformations already
observed in previous works, we found pseudo-phase transitions into novel phases
which are characterized by thermodynamically stable knots and showed that
$\Delta_p(-1.1)$ is a perfect topological order parameter for these
transitions. Moreover, at these topological transitions we found phase
coexistence between structured phases but, surprisingly, with equal total mean
energy in the two phases. The missing signal in the heat capacity implies a
pseudo-phase transition with a clear phase coexistence but without latent heat.
At the moment, investigating short single polymer chains in experiments seems
unrealistic, and it appears even more unrealistic to observe structural
properties such as knots. However, preparation and detection methods for single
polymers at surfaces have recently made quite an impressive
progress~\cite{Forster2014a,Forster2014}, so that one can now resolve polymers
of down to 20 monomers. A sensible next step could be hence an investigation of
the influence of an adsorbing surface on the occurrence of stable knotted
pseudo phases for a generic semi-flexible polymer.  Later on, this could be
extended to more realistic synthetic polymer models to guide experiments.

\begin{acknowledgments}
We would like to thank Johannes Zierenberg, Niklas Fricke, Stefan Schnabel, Jan
Meischner, and Martin Treff\-korn for many fruitful discussions. Computing time
provided by the John von Neumann Institute for Computing (NIC) under grant
No.~HLZ21 on the supercomputer JUROPA at J\"ulich Supercomputing Centre (JSC)
is gratefully acknowledged. This project has been funded by the European Union
and the Free State of Saxony through the ``Sächsische AufbauBank'' and by the
DFG (German Science Foundation) through SFB/TRR 102 (project B04) and the
Graduate School GSC 185 ``BuildMoNa''.
\end{acknowledgments}

\bibliography{paper}

\begin{thebibliography}{32}%
\makeatletter
\providecommand \@ifxundefined [1]{%
 \@ifx{#1\undefined}
}%
\providecommand \@ifnum [1]{%
 \ifnum #1\expandafter \@firstoftwo
 \else \expandafter \@secondoftwo
 \fi
}%
\providecommand \@ifx [1]{%
 \ifx #1\expandafter \@firstoftwo
 \else \expandafter \@secondoftwo
 \fi
}%
\providecommand \natexlab [1]{#1}%
\providecommand \enquote  [1]{``#1''}%
\providecommand \bibnamefont  [1]{#1}%
\providecommand \bibfnamefont [1]{#1}%
\providecommand \citenamefont [1]{#1}%
\providecommand \href@noop [0]{\@secondoftwo}%
\providecommand \href [0]{\begingroup \@sanitize@url \@href}%
\providecommand \@href[1]{\@@startlink{#1}\@@href}%
\providecommand \@@href[1]{\endgroup#1\@@endlink}%
\providecommand \@sanitize@url [0]{\catcode `\\12\catcode `\$12\catcode
  `\&12\catcode `\#12\catcode `\^12\catcode `\_12\catcode `\%12\relax}%
\providecommand \@@startlink[1]{}%
\providecommand \@@endlink[0]{}%
\providecommand \url  [0]{\begingroup\@sanitize@url \@url }%
\providecommand \@url [1]{\endgroup\@href {#1}{\urlprefix }}%
\providecommand \urlprefix  [0]{URL }%
\providecommand \Eprint [0]{\href }%
\providecommand \doibase [0]{http://dx.doi.org/}%
\providecommand \selectlanguage [0]{\@gobble}%
\providecommand \bibinfo  [0]{\@secondoftwo}%
\providecommand \bibfield  [0]{\@secondoftwo}%
\providecommand \translation [1]{[#1]}%
\providecommand \BibitemOpen [0]{}%
\providecommand \bibitemStop [0]{}%
\providecommand \bibitemNoStop [0]{.\EOS\space}%
\providecommand \EOS [0]{\spacefactor3000\relax}%
\providecommand \BibitemShut  [1]{\csname bibitem#1\endcsname}%
\let\auto@bib@innerbib\@empty
\bibitem [{\citenamefont {Frank-Kamenetskii}\ \emph {et~al.}(1975)\citenamefont
  {Frank-Kamenetskii}, \citenamefont {Lukashin},\ and\ \citenamefont
  {Vologodskii}}]{Frank-Kamenetskii1975}%
  \BibitemOpen
  \bibfield  {author} {\bibinfo {author} {\bibfnamefont {M.}~\bibnamefont
  {Frank-Kamenetskii}}, \bibinfo {author} {\bibfnamefont {A.}~\bibnamefont
  {Lukashin}}, \ and\ \bibinfo {author} {\bibfnamefont {A.}~\bibnamefont
  {Vologodskii}},\ }\href
  {http://www.nature.com/nature/journal/v258/n5534/abs/258398a0.html}
  {\bibfield  {journal} {\bibinfo  {journal} {Nature}\ }\textbf {\bibinfo
  {volume} {258}},\ \bibinfo {pages} {398} (\bibinfo {year}
  {1975})}\BibitemShut {NoStop}%
\bibitem [{\citenamefont {Mansfield}(1994)}]{Mansfield1994}%
  \BibitemOpen
  \bibfield  {author} {\bibinfo {author} {\bibfnamefont {M.}~\bibnamefont
  {Mansfield}},\ }\href {\doibase doi:10.1038/nsb0494-213} {\bibfield
  {journal} {\bibinfo  {journal} {Nat. Struct. Mol. Biol.}\ }\textbf {\bibinfo
  {volume} {1}},\ \bibinfo {pages} {213} (\bibinfo {year} {1994})}\BibitemShut
  {NoStop}%
\bibitem [{\citenamefont {Taylor}(2000)}]{Taylor2000}%
  \BibitemOpen
  \bibfield  {author} {\bibinfo {author} {\bibfnamefont {W.~R.}\ \bibnamefont
  {Taylor}},\ }\href {\doibase 10.1038/35022623} {\bibfield  {journal}
  {\bibinfo  {journal} {Nature}\ }\textbf {\bibinfo {volume} {406}},\ \bibinfo
  {pages} {916} (\bibinfo {year} {2000})}\BibitemShut {NoStop}%
\bibitem [{\citenamefont {Lua}\ and\ \citenamefont {Grosberg}(2006)}]{Lua2006}%
  \BibitemOpen
  \bibfield  {author} {\bibinfo {author} {\bibfnamefont {R.~C.}\ \bibnamefont
  {Lua}}\ and\ \bibinfo {author} {\bibfnamefont {A.~Y.}\ \bibnamefont
  {Grosberg}},\ }\href {\doibase 10.1371/journal.pcbi.0020045} {\bibfield
  {journal} {\bibinfo  {journal} {PLoS Comput. Biol.}\ }\textbf {\bibinfo
  {volume} {2}},\ \bibinfo {pages} {0350} (\bibinfo {year} {2006})}\BibitemShut
  {NoStop}%
\bibitem [{\citenamefont {Virnau}\ \emph {et~al.}(2006)\citenamefont {Virnau},
  \citenamefont {Mirny},\ and\ \citenamefont {Kardar}}]{Virnau2006}%
  \BibitemOpen
  \bibfield  {author} {\bibinfo {author} {\bibfnamefont {P.}~\bibnamefont
  {Virnau}}, \bibinfo {author} {\bibfnamefont {L.}~\bibnamefont {Mirny}}, \
  and\ \bibinfo {author} {\bibfnamefont {M.}~\bibnamefont {Kardar}},\ }\href
  {\doibase 10.1371/journal.pcbi.0020122} {\bibfield  {journal} {\bibinfo
  {journal} {PLoS Comput. Biol.}\ }\textbf {\bibinfo {volume} {2}},\ \bibinfo
  {pages} {1075} (\bibinfo {year} {2006})}\BibitemShut {NoStop}%
\bibitem [{\citenamefont {Berman}\ \emph {et~al.}(2000)\citenamefont {Berman},
  \citenamefont {Westbrook}, \citenamefont {Feng}, \citenamefont {Gilliland},
  \citenamefont {Bhat}, \citenamefont {Weissig}, \citenamefont {Shindyalov},\
  and\ \citenamefont {Bourne}}]{Berman2000}%
  \BibitemOpen
  \bibfield  {author} {\bibinfo {author} {\bibfnamefont {H.~M.}\ \bibnamefont
  {Berman}}, \bibinfo {author} {\bibfnamefont {J.}~\bibnamefont {Westbrook}},
  \bibinfo {author} {\bibfnamefont {Z.}~\bibnamefont {Feng}}, \bibinfo {author}
  {\bibfnamefont {Z.}~\bibnamefont {Gilliland}}, \bibinfo {author}
  {\bibfnamefont {T.~N.}\ \bibnamefont {Bhat}}, \bibinfo {author}
  {\bibfnamefont {H.}~\bibnamefont {Weissig}}, \bibinfo {author} {\bibfnamefont
  {I.~N.}\ \bibnamefont {Shindyalov}}, \ and\ \bibinfo {author} {\bibfnamefont
  {P.~E.}\ \bibnamefont {Bourne}},\ }\href {\doibase doi: 10.1093/nar/28.1.235}
  {\bibfield  {journal} {\bibinfo  {journal} {Nucleic Acids Res.}\ }\textbf
  {\bibinfo {volume} {28}},\ \bibinfo {pages} {235} (\bibinfo {year}
  {2000})}\BibitemShut {NoStop}%
\bibitem [{\citenamefont {W\"{u}st}\ \emph {et~al.}(2015)\citenamefont
  {W\"{u}st}, \citenamefont {Reith},\ and\ \citenamefont {Virnau}}]{Reith2015}%
  \BibitemOpen
  \bibfield  {author} {\bibinfo {author} {\bibfnamefont {T.}~\bibnamefont
  {W\"{u}st}}, \bibinfo {author} {\bibfnamefont {D.}~\bibnamefont {Reith}}, \
  and\ \bibinfo {author} {\bibfnamefont {P.}~\bibnamefont {Virnau}},\ }\href
  {\doibase 10.1103/PhysRevLett.114.028102} {\bibfield  {journal} {\bibinfo
  {journal} {Phys. Rev. Lett.}\ }\textbf {\bibinfo {volume} {114}},\ \bibinfo
  {pages} {028102} (\bibinfo {year} {2015})}\BibitemShut {NoStop}%
\bibitem [{\citenamefont {Koniaris}\ and\ \citenamefont
  {Muthukumar}(1991)}]{Koniaris1991b}%
  \BibitemOpen
  \bibfield  {author} {\bibinfo {author} {\bibfnamefont {K.}~\bibnamefont
  {Koniaris}}\ and\ \bibinfo {author} {\bibfnamefont {M.}~\bibnamefont
  {Muthukumar}},\ }\href {\doibase
  http://dx.doi.org/10.1103/PhysRevLett.66.2211} {\bibfield  {journal}
  {\bibinfo  {journal} {Phys. Rev. Lett.}\ }\textbf {\bibinfo {volume} {66}},\
  \bibinfo {pages} {2211} (\bibinfo {year} {1991})}\BibitemShut {NoStop}%
\bibitem [{\citenamefont {Deguchi}\ and\ \citenamefont
  {Tsurusaki}(1997)}]{Deguchi1997}%
  \BibitemOpen
  \bibfield  {author} {\bibinfo {author} {\bibfnamefont {T.}~\bibnamefont
  {Deguchi}}\ and\ \bibinfo {author} {\bibfnamefont {K.}~\bibnamefont
  {Tsurusaki}},\ }\href {\doibase 10.1103/PhysRevE.55.6245} {\bibfield
  {journal} {\bibinfo  {journal} {Phys. Rev. E}\ }\textbf {\bibinfo {volume}
  {55}},\ \bibinfo {pages} {6245} (\bibinfo {year} {1997})}\BibitemShut
  {NoStop}%
\bibitem [{\citenamefont {Virnau}\ \emph {et~al.}(2005)\citenamefont {Virnau},
  \citenamefont {Kantor},\ and\ \citenamefont {Kardar}}]{Virnau2005a}%
  \BibitemOpen
  \bibfield  {author} {\bibinfo {author} {\bibfnamefont {P.}~\bibnamefont
  {Virnau}}, \bibinfo {author} {\bibfnamefont {Y.}~\bibnamefont {Kantor}}, \
  and\ \bibinfo {author} {\bibfnamefont {M.}~\bibnamefont {Kardar}},\ }\href
  {\doibase 10.1021/ja052438a} {\bibfield  {journal} {\bibinfo  {journal} {J.
  Am. Chem. Soc.}\ }\textbf {\bibinfo {volume} {127}},\ \bibinfo {pages}
  {15102} (\bibinfo {year} {2005})}\BibitemShut {NoStop}%
\bibitem [{\citenamefont {Lua}\ \emph {et~al.}(2004)\citenamefont {Lua},
  \citenamefont {Borovinskiy},\ and\ \citenamefont {Grosberg}}]{Lua2004}%
  \BibitemOpen
  \bibfield  {author} {\bibinfo {author} {\bibfnamefont {R.}~\bibnamefont
  {Lua}}, \bibinfo {author} {\bibfnamefont {A.~L.}\ \bibnamefont
  {Borovinskiy}}, \ and\ \bibinfo {author} {\bibfnamefont {A.~Y.}\ \bibnamefont
  {Grosberg}},\ }\href {\doibase 10.1016/j.polymer.2003.10.073} {\bibfield
  {journal} {\bibinfo  {journal} {Polymer}\ }\textbf {\bibinfo {volume} {45}},\
  \bibinfo {pages} {717} (\bibinfo {year} {2004})}\BibitemShut {NoStop}%
\bibitem [{\citenamefont {Bastolla}\ and\ \citenamefont
  {Grassberger}(1997)}]{Bastolla1997}%
  \BibitemOpen
  \bibfield  {author} {\bibinfo {author} {\bibfnamefont {U.}~\bibnamefont
  {Bastolla}}\ and\ \bibinfo {author} {\bibfnamefont {P.}~\bibnamefont
  {Grassberger}},\ }\href {\doibase http://dx.doi.org/10.1007/BF02764222}
  {\bibfield  {journal} {\bibinfo  {journal} {J. Stat. Phys.}\ }\textbf
  {\bibinfo {volume} {89}},\ \bibinfo {pages} {1061} (\bibinfo {year}
  {1997})}\BibitemShut {NoStop}%
\bibitem [{\citenamefont {Krawczyk}\ \emph {et~al.}(2010)\citenamefont
  {Krawczyk}, \citenamefont {Owczarek},\ and\ \citenamefont
  {Prellberg}}]{Krawczyk2010}%
  \BibitemOpen
  \bibfield  {author} {\bibinfo {author} {\bibfnamefont {J.}~\bibnamefont
  {Krawczyk}}, \bibinfo {author} {\bibfnamefont {A.~L.}\ \bibnamefont
  {Owczarek}}, \ and\ \bibinfo {author} {\bibfnamefont {T.}~\bibnamefont
  {Prellberg}},\ }\href {\doibase 10.1016/j.physa.2009.12.012} {\bibfield
  {journal} {\bibinfo  {journal} {Physica A}\ }\textbf {\bibinfo {volume}
  {389}},\ \bibinfo {pages} {1619} (\bibinfo {year} {2010})}\BibitemShut
  {NoStop}%
\bibitem [{\citenamefont {Noguchi}\ and\ \citenamefont
  {Yoshikawa}(1997)}]{Noguchi1997}%
  \BibitemOpen
  \bibfield  {author} {\bibinfo {author} {\bibfnamefont {H.}~\bibnamefont
  {Noguchi}}\ and\ \bibinfo {author} {\bibfnamefont {K.}~\bibnamefont
  {Yoshikawa}},\ }\href@noop {} {\bibfield  {journal} {\bibinfo  {journal}
  {Chem. Phys. Lett.}\ }\textbf {\bibinfo {volume} {278}},\ \bibinfo {pages}
  {184} (\bibinfo {year} {1997})}\BibitemShut {NoStop}%
\bibitem [{\citenamefont {Seaton}\ \emph {et~al.}(2013)\citenamefont {Seaton},
  \citenamefont {Schnabel}, \citenamefont {Landau},\ and\ \citenamefont
  {Bachmann}}]{Seaton2013}%
  \BibitemOpen
  \bibfield  {author} {\bibinfo {author} {\bibfnamefont {D.~T.}\ \bibnamefont
  {Seaton}}, \bibinfo {author} {\bibfnamefont {S.}~\bibnamefont {Schnabel}},
  \bibinfo {author} {\bibfnamefont {D.~P.}\ \bibnamefont {Landau}}, \ and\
  \bibinfo {author} {\bibfnamefont {M.}~\bibnamefont {Bachmann}},\ }\href
  {\doibase 10.1103/PhysRevLett.110.028103} {\bibfield  {journal} {\bibinfo
  {journal} {Phys. Rev. Lett.}\ }\textbf {\bibinfo {volume} {110}},\ \bibinfo
  {pages} {028103} (\bibinfo {year} {2013})}\BibitemShut {NoStop}%
\bibitem [{\citenamefont {Stillinger}\ \emph {et~al.}(1993)\citenamefont
  {Stillinger}, \citenamefont {Head-Gordon},\ and\ \citenamefont
  {Hirshfeld}}]{Stillinger1993}%
  \BibitemOpen
  \bibfield  {author} {\bibinfo {author} {\bibfnamefont {F.~H.}\ \bibnamefont
  {Stillinger}}, \bibinfo {author} {\bibfnamefont {T.}~\bibnamefont
  {Head-Gordon}}, \ and\ \bibinfo {author} {\bibfnamefont {C.~L.}\ \bibnamefont
  {Hirshfeld}},\ }\href {http://pre.aps.org/abstract/PRE/v48/i2/p1469\_1}
  {\bibfield  {journal} {\bibinfo  {journal} {Phys. Rev. E}\ }\textbf {\bibinfo
  {volume} {48}},\ \bibinfo {pages} {1469} (\bibinfo {year}
  {1993})}\BibitemShut {NoStop}%
\bibitem [{\citenamefont {Irb\"{a}ck}\ \emph {et~al.}(1997)\citenamefont
  {Irb\"{a}ck}, \citenamefont {Peterson}, \citenamefont {Potthast},\ and\
  \citenamefont {Sommelius}}]{Irback1997}%
  \BibitemOpen
  \bibfield  {author} {\bibinfo {author} {\bibfnamefont {A.}~\bibnamefont
  {Irb\"{a}ck}}, \bibinfo {author} {\bibfnamefont {C.}~\bibnamefont
  {Peterson}}, \bibinfo {author} {\bibfnamefont {F.}~\bibnamefont {Potthast}},
  \ and\ \bibinfo {author} {\bibfnamefont {O.}~\bibnamefont {Sommelius}},\
  }\href {\doibase 10.1063/1.474357} {\bibfield  {journal} {\bibinfo  {journal}
  {J. Chem. Phys.}\ }\textbf {\bibinfo {volume} {107}},\ \bibinfo {pages} {273}
  (\bibinfo {year} {1997})}\BibitemShut {NoStop}%
\bibitem [{\citenamefont {Bachmann}\ \emph {et~al.}(2005)\citenamefont
  {Bachmann}, \citenamefont {Arkin},\ and\ \citenamefont
  {Janke}}]{Bachmann2005}%
  \BibitemOpen
  \bibfield  {author} {\bibinfo {author} {\bibfnamefont {M.}~\bibnamefont
  {Bachmann}}, \bibinfo {author} {\bibfnamefont {H.}~\bibnamefont {Arkin}}, \
  and\ \bibinfo {author} {\bibfnamefont {W.}~\bibnamefont {Janke}},\ }\href
  {\doibase 10.1103/PhysRevE.71.031906} {\bibfield  {journal} {\bibinfo
  {journal} {Phys. Rev. E}\ }\textbf {\bibinfo {volume} {71}},\ \bibinfo
  {pages} {031906} (\bibinfo {year} {2005})}\BibitemShut {NoStop}%
\bibitem [{\citenamefont {Rubinstein}\ and\ \citenamefont
  {Colby}(2003)}]{Rubinstein2003}%
  \BibitemOpen
  \bibfield  {author} {\bibinfo {author} {\bibfnamefont {M.}~\bibnamefont
  {Rubinstein}}\ and\ \bibinfo {author} {\bibfnamefont {R.~H.}\ \bibnamefont
  {Colby}},\ }\href@noop {} {\emph {\bibinfo {title} {{Polymer Physics}}}}\
  (\bibinfo  {publisher} {Oxford University Press},\ \bibinfo {address}
  {Oxford},\ \bibinfo {year} {2003})\ p.\ \bibinfo {pages} {456}\BibitemShut
  {NoStop}%
\bibitem [{\citenamefont {Zierenberg}\ \emph {et~al.}(2013)\citenamefont
  {Zierenberg}, \citenamefont {Marenz},\ and\ \citenamefont
  {Janke}}]{Zierenberg2013}%
  \BibitemOpen
  \bibfield  {author} {\bibinfo {author} {\bibfnamefont {J.}~\bibnamefont
  {Zierenberg}}, \bibinfo {author} {\bibfnamefont {M.}~\bibnamefont {Marenz}},
  \ and\ \bibinfo {author} {\bibfnamefont {W.}~\bibnamefont {Janke}},\ }\href
  {\doibase 10.1016/j.cpc.2012.12.006} {\bibfield  {journal} {\bibinfo
  {journal} {Comput. Phys. Commun.}\ }\textbf {\bibinfo {volume} {184}},\
  \bibinfo {pages} {1155} (\bibinfo {year} {2013})}\BibitemShut {NoStop}%
\bibitem [{\citenamefont {Berg}\ and\ \citenamefont
  {Neuhaus}(1991)}]{Berg1991}%
  \BibitemOpen
  \bibfield  {author} {\bibinfo {author} {\bibfnamefont {B.~A.}\ \bibnamefont
  {Berg}}\ and\ \bibinfo {author} {\bibfnamefont {T.}~\bibnamefont {Neuhaus}},\
  }\href {\doibase 10.1016/0370-2693(91)91256-U} {\bibfield  {journal}
  {\bibinfo  {journal} {Phys. Lett. B}\ }\textbf {\bibinfo {volume} {267}},\
  \bibinfo {pages} {249} (\bibinfo {year} {1991})}\BibitemShut {NoStop}%
\bibitem [{\citenamefont {Berg}\ and\ \citenamefont
  {Neuhaus}(1992)}]{Berg1992}%
  \BibitemOpen
  \bibfield  {author} {\bibinfo {author} {\bibfnamefont {B.~A.}\ \bibnamefont
  {Berg}}\ and\ \bibinfo {author} {\bibfnamefont {T.}~\bibnamefont {Neuhaus}},\
  }\href {\doibase DOI: http://dx.doi.org/10.1103/PhysRevLett.68.9} {\bibfield
  {journal} {\bibinfo  {journal} {Phys. Rev. Lett.}\ }\textbf {\bibinfo
  {volume} {68}},\ \bibinfo {pages} {9} (\bibinfo {year} {1992})}\BibitemShut
  {NoStop}%
\bibitem [{\citenamefont {Hukushima}\ and\ \citenamefont
  {Nemoto}(1996)}]{Hukushima1995}%
  \BibitemOpen
  \bibfield  {author} {\bibinfo {author} {\bibfnamefont {K.}~\bibnamefont
  {Hukushima}}\ and\ \bibinfo {author} {\bibfnamefont {K.}~\bibnamefont
  {Nemoto}},\ }\href {\doibase 10.1143/JPSJ.65.1604} {\bibfield  {journal}
  {\bibinfo  {journal} {J. Phys. Soc. Japan}\ }\textbf {\bibinfo {volume}
  {65}},\ \bibinfo {pages} {1604} (\bibinfo {year} {1996})}\BibitemShut
  {NoStop}%
\bibitem [{\citenamefont {Ferrenberg}\ and\ \citenamefont
  {Swendsen}(1988)}]{Ferrenberg1988}%
  \BibitemOpen
  \bibfield  {author} {\bibinfo {author} {\bibfnamefont {A.~M.}\ \bibnamefont
  {Ferrenberg}}\ and\ \bibinfo {author} {\bibfnamefont {R.~H.}\ \bibnamefont
  {Swendsen}},\ }\href {http://link.aps.org/doi/10.1103/PhysRevLett.61.2635}
  {\bibfield  {journal} {\bibinfo  {journal} {Phys. Rev. Lett.}\ }\textbf
  {\bibinfo {volume} {61}},\ \bibinfo {pages} {2635} (\bibinfo {year}
  {1988})}\BibitemShut {NoStop}%
\bibitem [{\citenamefont {Ferrenberg}\ and\ \citenamefont
  {Swendsen}(1989)}]{Ferrenberg1989}%
  \BibitemOpen
  \bibfield  {author} {\bibinfo {author} {\bibfnamefont {A.~M.}\ \bibnamefont
  {Ferrenberg}}\ and\ \bibinfo {author} {\bibfnamefont {R.~H.}\ \bibnamefont
  {Swendsen}},\ }\href@noop {} {\bibfield  {journal} {\bibinfo  {journal}
  {Phys. Rev. Lett.}\ }\textbf {\bibinfo {volume} {63}},\ \bibinfo {pages}
  {1195} (\bibinfo {year} {1989})}\BibitemShut {NoStop}%
\bibitem [{\citenamefont {Junghans}\ \emph {et~al.}(2006)\citenamefont
  {Junghans}, \citenamefont {Bachmann},\ and\ \citenamefont
  {Janke}}]{Junghans2006}%
  \BibitemOpen
  \bibfield  {author} {\bibinfo {author} {\bibfnamefont {C.}~\bibnamefont
  {Junghans}}, \bibinfo {author} {\bibfnamefont {M.}~\bibnamefont {Bachmann}},
  \ and\ \bibinfo {author} {\bibfnamefont {W.}~\bibnamefont {Janke}},\ }\href
  {\doibase 10.1103/PhysRevLett.97.218103} {\bibfield  {journal} {\bibinfo
  {journal} {Phys. Rev. Lett.}\ }\textbf {\bibinfo {volume} {97}},\ \bibinfo
  {pages} {218103} (\bibinfo {year} {2006})}\BibitemShut {NoStop}%
\bibitem [{\citenamefont {Kauffmann}(1991)}]{Kauffmann1991}%
  \BibitemOpen
  \bibfield  {author} {\bibinfo {author} {\bibfnamefont {L.~H.}\ \bibnamefont
  {Kauffmann}},\ }\href@noop {} {\emph {\bibinfo {title} {{Knots and
  Physics}}}},\ \bibinfo {edition} {2nd}\ ed.\ (\bibinfo  {publisher} {World
  Scientific},\ \bibinfo {address} {Singapore},\ \bibinfo {year}
  {1991})\BibitemShut {NoStop}%
\bibitem [{\citenamefont {Virnau}(2010)}]{Virnau2010}%
  \BibitemOpen
  \bibfield  {author} {\bibinfo {author} {\bibfnamefont {P.}~\bibnamefont
  {Virnau}},\ }\href {\doibase 10.1016/j.phpro.2010.09.036} {\bibfield
  {journal} {\bibinfo  {journal} {Physics Procedia}\ }\textbf {\bibinfo
  {volume} {6}},\ \bibinfo {pages} {117} (\bibinfo {year} {2010})}\BibitemShut
  {NoStop}%
\bibitem [{Note1()}]{Note1}%
  \BibitemOpen
  \bibinfo {note} {Not shown here, because the details of the more complicated
  pseudo-phase diagram do not contribute to the understanding of the basic
  mechanism.}\BibitemShut {Stop}%
\bibitem [{Note2()}]{Note2}%
  \BibitemOpen
  \bibinfo {note} {This topological change also explains why we need PMUCA+RE
  or 2D-RE to overcome the topological barrier.}\BibitemShut {Stop}%
\bibitem [{\citenamefont {F\"{o}rster}\ and\ \citenamefont
  {Widdra}(2014)}]{Forster2014a}%
  \BibitemOpen
  \bibfield  {author} {\bibinfo {author} {\bibfnamefont {S.}~\bibnamefont
  {F\"{o}rster}}\ and\ \bibinfo {author} {\bibfnamefont {W.}~\bibnamefont
  {Widdra}},\ }\href {\doibase 10.1063/1.4891929} {\bibfield  {journal}
  {\bibinfo  {journal} {J. Chem. Phys.}\ }\textbf {\bibinfo {volume} {141}},\
  \bibinfo {pages} {054713} (\bibinfo {year} {2014})}\BibitemShut {NoStop}%
\bibitem [{\citenamefont {F\"{o}rster}\ \emph {et~al.}(2014)\citenamefont
  {F\"{o}rster}, \citenamefont {Kohl}, \citenamefont {Ivanov}, \citenamefont
  {Gross}, \citenamefont {Widdra},\ and\ \citenamefont {Janke}}]{Forster2014}%
  \BibitemOpen
  \bibfield  {author} {\bibinfo {author} {\bibfnamefont {S.}~\bibnamefont
  {F\"{o}rster}}, \bibinfo {author} {\bibfnamefont {E.}~\bibnamefont {Kohl}},
  \bibinfo {author} {\bibfnamefont {M.}~\bibnamefont {Ivanov}}, \bibinfo
  {author} {\bibfnamefont {J.}~\bibnamefont {Gross}}, \bibinfo {author}
  {\bibfnamefont {W.}~\bibnamefont {Widdra}}, \ and\ \bibinfo {author}
  {\bibfnamefont {W.}~\bibnamefont {Janke}},\ }\href {\doibase
  10.1063/1.4898382} {\bibfield  {journal} {\bibinfo  {journal} {J. Chem.
  Phys.}\ }\textbf {\bibinfo {volume} {141}},\ \bibinfo {pages} {164701}
  (\bibinfo {year} {2014})}\BibitemShut {NoStop}%
\end{thebibliography}%

\end{document}